\documentclass[12pt]{article}
\newcommand{\nc}{\newcommand}
\nc{\fh}{\hat{f}}
\nc{\muh}{\hat{\mu}}
\nc{\nuh}{\hat{\nu}}
\nc{\kh}{\hat{\kappa}}
\nc{\hh}{\hat{h}}
\nc{\gh}{\hat{g}}
\nc{\bib}{\bibitem}
\nc{\al}{\alpha}
\nc{\g}{\gamma}
\nc{\G}{\Gamma}
\nc{\D}{\Delta}
\nc{\eps}{\epsilon}
\nc{\la}{\lambda}
\nc{\La}{\Lambda}
\nc{\var}{\varphi}
\nc{\cg}{{\cal G}}
\nc{\pa}{\partial}
\nc{\nn}{\nonumber \\ }
\nc{\hf}{\frac{1}{2}}  
\nc{\dz}{\frac{dz}{2\pi i}}
\nc{\bin}[2]{\left (\begin{array}{c} {#1}\\ {#2} \end{array}\right )}
\nc{\ben}{\begin{equation}}
\nc{\een}{\end{equation}}
\nc{\bea}{\begin{eqnarray}}
\nc{\eea}{\end{eqnarray}}
\nc{\bra}[1]{\langle {#1}|}
\nc{\ket}[1]{|{#1}\rangle}
\newcommand{\Z}{\mbox{$Z\hspace{-2mm}Z$}}
\nc{\C}{\mbox{\hspace{1.24mm}\rule{0.2mm}{2.5mm}\hspace{-2.7mm} C}}
\nc{\Nat}{\mbox{\hspace{.04mm}\rule{0.2mm}{2.8mm}\hspace{-1.5mm} N}}

\nc{\HH}{\mbox{\hspace{.04mm}\rule{0.2mm}{2.8mm}\hspace{-1.5mm} H}}

\def\vvdots{\mathinner{\mkern1mu\raise1pt\vbox{\kern7pt\hbox{.}}\mkern2mu
 \raise4pt\hbox{.}\mkern2mu\raise7pt\hbox{.}\mkern1mu}}

\begin{document}

\topmargin -5mm
\oddsidemargin 5mm

\begin{titlepage}
\setcounter{page}{0}

\vspace{8mm}
\begin{center}
{\huge Note on SLE and logarithmic CFT}

\vspace{15mm}
{\Large J{\o}rgen Rasmussen}\\[.3cm] 
{\em Centre de recherches math\'ematiques, Universit\'e de Montr\'eal}\\ 
{\em Case postale 6128, 
succursale centre-ville, Montr\'eal, Qc, Canada H3C 3J7}\\[.3cm]
rasmusse@crm.umontreal.ca

\end{center}

\vspace{10mm}
\centerline{{\bf{Abstract}}}
\vskip.4cm
\noindent
It is discussed how stochastic evolutions may be linked to 
logarithmic conformal field theory. This introduces an extension of
the stochastic L\"owner evolutions.
Based on the existence of a logarithmic null vector in
an indecomposable highest-weight module of the Virasoro algebra,
the representation theory of the logarithmic conformal field theory
is related to entities conserved in mean under the stochastic process.
\\[.5cm]
{\bf Keywords:} Logarithmic conformal field theory (CFT), stochastic 
L\"owner evolution (SLE). 
\end{titlepage}
\newpage
\renewcommand{\thefootnote}{\arabic{footnote}}
\setcounter{footnote}{0}

\section{Introduction}

Stochastic L\"owner evolutions (SLEs) have been introduced by Schramm
\cite{Sch} and further developed in \cite{LSW,RS} as a mathematically
rigorous way of describing certain two-dimensional systems at criticality.
The method involves the study of stochastic evolutions of conformal maps,
and puts into a new mathematical framework what conformal field theory (CFT)
has been addressing for decades.
Applications as well as formal properties and generalizations of SLE
have since been investigated from various points of view.
Reviews on SLE may be found in \cite{Law,KN}.

The present work suggests that also logarithmic CFT (LCFT) 
may be linked to SLE. We refer to 
\cite{Flo,Gab,Nic} for recent reviews on LCFT, and to 
\cite{Gur} for the first systematic study of that subject.
The basic observation made here is that the approach of Bauer and
Bernard \cite{BB} (see also \cite{FW}) may be extended from
ordinary CFT to LCFT
where some primary fields are accompanied by Jordan-cell partners
with new transformation properties. 
Other extensions of \cite{BB} may be found in 
\cite{LR,Ras,NR,su2sle}. CFT and SLE-type growth processes in
smaller regions of the complex plane than in ordinary chordal SLE are
studied in \cite{LR}. The references \cite{Ras,NR} discuss how
stochastic evolutions in superspace may be linked to superconformal 
field theory, whereas \cite{su2sle} concerns the elevation to 
$SU(2)$ Wess-Zumino-Witten models with particular emphasis on
the Sugawara construction and the Knizhnik-Zamolodchikov equations.

The initial relationship between stochastic evolutions and LCFT 
is mediated by
stochastic differential equations and random walks on the Virasoro group.
A particular scenario is outlined that extends SLE from being described
by one stochastic differential equation to a system evolving
according to a pair of coupled stochastic differential equations. 

The relationship can be made more direct by establishing
a connection between the representation theory of the LCFT
and entities conserved in mean under the stochastic
process. This is based on the existence of a level-two
null vector in the associated indecomposable
highest-weight module of the Virasoro algebra. 
Particular attention is paid to the module with
conformal weight $\D=1/4$ as it has a so-called logarithmic
null vector when the central charge is $c=1$ \cite{Flo-9707}. 
The description relies on the introduction of a nilpotent parameter
\cite{Flo-9707,MRS}. This suggests extending the regime of SLE from 
the complex plane to stochastic evolutions in the space obtained by 
augmenting the complex plane by this parameter.
Alternatively, and perhaps more straightforwardly,
one may interpret the result as two coupled 
stochastic evolutions. One of them is SLE$_{\kappa=4}$
and corresponds to the SLE 
phase transition between simple paths (for $0\leq\kappa\leq4$) and
self-intersecting, or rather self-touching, paths (for $4<\kappa<8$)
defined by the so-called SLE trace. SLE$_4$ also corresponds to the
self-dual point of a duality between
SLE$_\kappa$ and SLE$_{16/\kappa}$ where one model is
linked to the description of the boundary of the 
SLE hull of the other model \cite{Dup,Dub}.
A discussion of SLE$_4$ itself may be found in \cite{SS}.

The link between SLE and LCFT is considered in section 2, whereas
section 3 contains some concluding remarks.

\section{Linking SLE to LCFT}

\subsection{Rudiments of LCFT}

A conformal Jordan cell of rank two consists of two fields:
a primary field, $\Phi$, of conformal weight $\D$ and
its non-primary partner, $\Psi$, on which the Virasoro algebra generated 
by $\{L_n\}$
does not act diagonally. With a conventional relative normalization
of the fields, we have
\bea
 T(z)\Phi(w)&=&\frac{\D\Phi(w)}{(z-w)^2}+\frac{\pa\Phi(w)}{z-w}\nn
 T(z)\Psi(w)&=&\frac{\D\Psi(w)+\Phi(w)}{(z-w)^2}
   +\frac{\pa\Psi(w)}{z-w}
\label{T}
\eea
where $T$ is the energy-momentum tensor whose mode expansion
is given in terms of $\{L_n\}$. In terms of these modes, (\ref{T}) reads
\bea
 \left[L_n,\Phi(z)\right]&=&\left(z^{n+1}\pa_z+\D(n+1)z^n\right)\Phi(z)\nn
 \left[L_n,\Psi(z)\right]&=&\left(z^{n+1}\pa_z+\D(n+1)z^n\right)\Psi(z)
  +(n+1)z^n\Phi(z)
\label{L}
\eea
The two-point functions are then of the form
\ben
 \langle\Phi(z)\Phi(w)\rangle\ =\ 0,\ \ \ \ 
 \langle\Phi(z)\Psi(w)\rangle\ =\ \frac{A}{(z-w)^{2\D}},\ \ \ \ 
 \langle\Psi(z)\Psi(w)\rangle\ =\ \frac{B-2A\ln(z-w)}{(z-w)^{2\D}}
\label{two}
\een
with structure constants $A$ and $B$.

The two fields transform as
\bea
 \Phi(z)&\rightarrow&(f'(z))^\D\Phi(f(z))\nn
 \Psi(z)&\rightarrow&(f'(z))^\D\left\{\Psi(f(z))+\ln(f'(z))\Phi(f(z))\right\}
\label{trans}
\eea
where a prime indicates a derivative with respect to the complex argument,
that is, $f'(z)=\pa_z f(z)$.
It was suggested by Flohr \cite{Flo-9707} to describe these transformations
in a unified way by introducing a nilpotent, yet even, parameter
$\theta$ satisfying $\theta^2=0$. We shall follow this idea here,
though use an approach closer to the one employed in \cite{MRS}.
We thus define the field
\ben
 \Upsilon(z,\theta)\ =\ \Phi(z)+\theta\Psi(z)
\label{upsilon}
\een
which is seen to be 'primary' of conformal weight $\D+\theta$ as
it transforms like
\ben
 \Upsilon(z,\theta)\ \rightarrow\ (f'(z))^{\D+\theta}\Upsilon(f(z),\theta)
\label{transup}
\een
This follows from the expansion
\ben
 x^\eps\ =\ e^{\eps\ln(x)}\ =\ 1+\eps\ln(x)+{\cal O}(\eps^2)
\label{x}
\een
which is exact for the nilpotent parameter $\eps=\theta$. 
The commutators (\ref{L}) are now replaced by
\ben
 \left[L_n,\Upsilon(z,\theta)\right]\ =\ 
  \left(z^{n+1}\pa_z+(\D+\theta)(n+1)z^n\right)\Upsilon(z,\theta)
\label{Lup}
\een

\subsection{Stochastic differentials}

We now consider the stochastic or Ito differential
\ben
 \cg^{-1}_t(\tau)d\cg_t(\tau)
  \ =\ \al(\tau) dt+\beta(\tau) dB_t ,\ \ \ \ \ \ \ \ \cg_0(\tau)=1
\label{cg}
\een
where we can think of $\cg_t(\tau)$ as a $\tau$-dependent
random walk on the Virasoro group. The parameter $\tau$ is
nilpotent, satisfying $\tau^2=0$, and thus of the same nature as 
$\theta$ introduced above. The reason for letting $\cg_t$
depend on $\tau$ will become clear when discussing null
vectors below.
We are confining ourselves to one-dimensional Brownian motion,
$B_t$, with $B_0=0$, while
$\al$ and $\beta$ generically are non-commutative
expressions in the generators of the Virasoro algebra.
It follows from $d(\cg^{-1}_t\cg_t)=0$ that the Ito differential
of the inverse element is given by
\ben
 d(\cg^{-1}_t)\cg_t\ =\ (-\al+\beta^2)dt-\beta dB_t
\label{cg-1}
\een

We are interested in the transformations of $\Upsilon$
generated by $\cg_t$. An ordinary conformal transformation
(\ref{transup}) generated by $\cg_t$ acts on $\Upsilon$ like
\ben
 \cg^{-1}_t\Upsilon(z,\theta)\cg_t\ 
  =\ \left(f_t'(z)\right)^{\D+\theta} \Upsilon(f_t(z),\theta)
\label{Gup}
\een
where $f_t(z)$ is a stochastic function of $z$.
This, however, requires that $\cg_t$ depends trivially on $\tau$.
We shall therefore discuss the extension of (\ref{Gup}) where the function
$f_t$ is allowed to depend on $\tau$. It is still conformal with respect to
$z$ while the otherwise only {\em complex} expansion coefficients 
now may depend on $\tau$ as well. This modifies the transformation rule 
(\ref{Gup}) as it becomes
\ben
 \cg^{-1}_t(\tau)\Upsilon(z,\theta)\cg_t(\tau)\ 
  =\ \left(f_t'(z,\tau)\right)^{\D+\theta} \Upsilon(f_t(z,\tau),\theta)
\label{Gupt}
\een
The function $f_t$ may be expanded as
\ben
 f_t(z,\tau)\ =\ h_t(z)+\tau \hh_t(z)
\label{hq}
\een
with Ito differential
\ben
 df_t(z,\tau)\ =\ \mu_t(z,\tau)dt+\nu_t(z,\tau)dB_t,\ \ \ \ \ \ \ 
   f_0(z,\tau)=z
\label{munut}
\een
To relax the notation,
the subscript $t$ will occasionally be suppressed below.
A goal is to compute the Ito differential of
both sides of (\ref{Gupt}) and thereby relate the
stochastic differential equations of $\cg_t(\tau)$ and $f_t(z,\tau)$.
It is recalled that in Ito calculus $(dB_t)^2=dt$ whereas 
$dtdB_t=(dt)^2=0$.

The Ito differential of the left-hand side of (\ref{Gupt}) reads
\bea
 d\left(\cg^{-1}_t\Upsilon\cg_t\right)&=&d(\cg^{-1}_t)\Upsilon\cg_t+
  \cg^{-1}_t\Upsilon d\cg_t+d(\cg^{-1}_t)\Upsilon d\cg_t\nn
 &=&\left(-\left[\al_0,\cg^{-1}_t\Upsilon\cg_t\right]
  +\hf\left[\beta,\left[\beta,\cg^{-1}_t\Upsilon\cg_t\right]\right]\right)dt\nn
  && -\left[\beta,\cg^{-1}_t\Upsilon\cg_t\right]dB_t
\eea
where $\al_0=\al-\hf\beta^2$.
To facilitate the comparison we should 
express the differentials in the {\em adjoint} representation
of the Virasoro algebra only. That is, 
$\al_0$ and $\beta$ must be linear in the generators:
\ben
 \al_0(\tau)\ =\ \sum_{n\in\Z}a_n(\tau)L_n,\ \ \ \ \ \ \ \ \ \ 
  \beta(\tau)\ =\ \sum_{n\in\Z}b_n(\tau)L_n
\label{al0}
\een
To avoid questions of convergence, we shall consider only 
finite sums, and we find
\bea
 d\left(\cg^{-1}_t\Upsilon(z)\cg_t\right)
 &=&\left(f'\right)^{\D+\theta}\left(-\left[\al_0,\Upsilon(f)\right]
   +\hf\left[\beta,\left[\beta,\Upsilon(f)
  \right]\right]\right)dt\nn
 &-&(f')^{\D+\theta}\sum_nb_n
   \{f^{n+1}\pa_f+(\D+\theta)(n+1)f^n\}\Upsilon(f)dB_t
\label{Dal0}
\eea
where the further evaluation is postponed till the result of
a comparison of the $dB_t$ terms can be used. 

The Ito differential of the right-hand side of (\ref{Gupt})
reads 
\bea
  d\left(\left(f'\right)^{\D+\theta} \Upsilon(f)\right)&=&
   (f')^{\D+\theta-2}\left(\hf(f')^2\nu^2\pa_f^2
    +\{(f')^2\mu+(\D+\theta)f'\nu\nu'\}\pa_f\right.\nn
  &&\left.+(\D+\theta)f'\mu'+\hf(\D+\theta)(\D+\theta-1)(\nu')^2\right)
   \Upsilon(f) dt\nn
  &+&(f')^{\D+\theta-1}\{f'\nu\pa_f+(\D+\theta)\nu'\}
   \Upsilon(f)dB_t
\label{drhs}
\eea
It follows from a comparison of the $dB_t$ terms that 
\ben
 f'[\beta,\Upsilon]=-\{f'\nu\pa_f+(\D+\theta)\nu'\}\Upsilon(f)
\label{betaup}
\een
and hence
\ben
 \nu_t\ =\ -\sum_{n}b_nf_t^{n+1}
\label{nu}
\een
The relation (\ref{betaup}) also allows us to continue the evaluation
of (\ref{Dal0}), and we find that
\ben
 \mu_t\ =\ -\sum_na_nf_t^{n+1}+\hf\nu_t\pa_{f_t}\nu_t
\label{mu}
\een
This is easily expressed in terms of $f_t$ only,
using (\ref{nu}), though the result is less compact than 
(\ref{mu}). Note that we get the same 
solution for $\mu_t$ and $\nu_t$ when considering an
ordinary primary field instead of the logarithmic
one in (\ref{Gupt}) (or the simpler one in (\ref{Gup})). 
This illustrates that the evaluation is {\em independent} 
of the parameter $\theta$ used in the construction of the
logarithmic field $\Upsilon(z,\theta)$. This is actually
required for the link to be universal and not depend 
explicitly on the field in question.

To summarize, we have found that the construction
above establishes a general but formal link between a class
of stochastic evolutions and LCFT: the Ito
differentials (\ref{munut}) describing the evolution of
the $\tau$-dependent conformal maps are expressed in 
terms of the parameters
of the random walk on the Virasoro group (\ref{cg}) and
(\ref{al0}), and this has been achieved via (\ref{Gupt}).
The links (\ref{nu}) and (\ref{mu}) may of course be expanded
with respect to $\tau$. The result of doing this does not seem to
be enlightening in its general form, and will therefore not be
presented here. An example follows below, though.

The construction and link above pertain to LCFT
based on rank-two Jordan cells. We believe, though, 
that the extension to higher rank \cite{RAK} is straightforward.

The scenario resembling SLE 
is based on the simple situation where $a_{-2}=-2$ and
$b_{-1}=\sqrt{k(\tau)}$ are the only non-vanishing coefficients:
\ben
 \cg^{-1}_t(\tau)d\cg_t(\tau)\ =\ \left(-2L_{-2}
   +\frac{k(\tau)}{2}L_{-1}^2\right)dt
   +\sqrt{k(\tau)}dB_t,\ \ \ \ \ \ \cg_0(\tau)=1
\label{SLEG}
\een
Here we have introduced the $\tau$-dependent parameter
\ben
 k(\tau)\ =\ \kappa+\tau\kh
\label{kappa}
\een
The Ito differential of the associated $\tau$-dependent
conformal map reads
\ben
 df_t(z,\tau)\ =\ \frac{2}{f_t(z,\tau)}dt-\sqrt{k(\tau)}
  dB_t,\ \ \ \ \ \ \ \ \ f_0(z,\tau)=z
\label{SLEf}
\een
Expanding this with respect to $\tau$ as in (\ref{hq}) 
leads to
\bea
 dh_t(z)&=&\frac{2}{h_t(z)}dt-\sqrt{\kappa}dB_t ,\ \ \ \ \ h_0(z)=z     \nn
 d\hh_t(z)&=&-\frac{2\hh_t(z)}{h_t^2(z)}dt-\frac{\kh}{2\sqrt{\kappa}}dB_t,
  \ \ \ \ \ \hh_0(z)=0
\label{dhdh}
\eea
In terms of the stochastic function $g_t(z)=h_t(z)+\sqrt{\kappa}B_t$,
the first stochastic differential equation in (\ref{dhdh}) 
may be expressed as the celebrated L\"owner's differential equation
with Brownian motion as driving function or potential, also known
as the SLE$_\kappa$ differential equation:
\ben
 \pa_t g_t(z)\ =\ \frac{2}{g_t(z)-\sqrt{\kappa}B_t},\ \ \ \ \ \ \ \ \ g_0(z)=z
\label{SLEg}
\een
This is now coupled to the differential equation
\ben
 \pa_t \gh_t(z)\ =\ \frac{-2\gh_t(z)+\frac{\kh}{\sqrt{\kappa}}B_t}{
   (g_t(z)-\sqrt{\kappa}B_t)^2},\ \ \ \ \ \ \ \ \ \gh_0(z)=0
\label{slealt}
\een
where we have introduced 
$\gh_t(z)=\hh_t(z)+\frac{\kh}{2\sqrt{\kappa}}B_t$.
Ordinary SLE is recovered by disregarding the additional
equation (\ref{slealt}) which is possible since (\ref{SLEg})
is independent of the function $\gh_t(z)$.
This essentially corresponds to omitting the dependence on
the nilpotent parameters altogether, which naturally reduces
the considerations to ordinary CFT and SLE.

\subsection{Expectation values and logarithmic null vectors}

To obtain a more direct relationship between SLE and LCFT, 
we should link the representation theory of the conformal algebra,
through the construction of logarithmic null vectors, to entities
conserved in mean under the stochastic process. 
{}From the group differential (\ref{cg})
it follows that the time evolution of the
expectation value of $\cg_t(\tau)\ket{\Delta+\theta}$ is given by
\ben
 \pa_t{\bf E}[\cg_t(\tau)\ket{\Delta+\theta}]\ =\ 
  {\bf E}[\cg_t(\tau)\left(\al_0(\tau)+
   \hf \beta^2(\tau)\right)\ket{\Delta+\theta}]
\label{EG}
\een
Here we have introduced the notation
\ben
 \ket{\D+\theta}\ =\ \ket{\Phi}+\theta\ket{\Psi}
\label{Dtheta}
\een
where (cf. (\ref{T}) and (\ref{L}))
\ben
 L_0\ket{\Phi}\ =\ \D\ket{\Phi},\ \ \ \ \ \ \ \ L_0\ket{\Psi}\ =\ 
  \D\ket{\Psi}+\ket{\Phi}
\label{L0}
\een
such that $L_0\ket{\D+\theta}=(\D+\theta)\ket{\D+\theta}$,
and $L_{n>0}\ket{\D+\theta}=0$.
We should thus look for processes allowing us to put
\ben
 \left(\al_0(\tau)+\hf \beta^2(\tau)\right)\ket{\D+\theta}\ \simeq \ 0
\label{0}
\een
in the representation theory. The expression
$\cg_t(\tau)\ket{\D+\theta}$ is then a so-called martingale of the 
stochastic process $\cg_t(\tau)$.

Before doing that, let us indicate how time evolutions of some
general expectation values may be evaluated.
This extends one of the main results in \cite{BB} on
ordinary SLE, and follows from the extension to the
graded case discussed in \cite{Ras,NR}.
In this regard, observables of the process ${\cal G}_t$
are thought of as functions of ${\cal G}_t$. On such a function
$F$, admitting a 'sufficiently convergent' Laurent expansion,
we introduce the action of the vector field $\nabla_n$ as
\ben
 (\nabla_nF)({\cal G}_t)\ =\ \frac{d}{du}F({\cal G}_t
  e^{uL_n})|_{u=0}
\label{nabla}
\een
Referring to the notation in (\ref{al0}), we then have
\ben
  \pa_t{\bf E}[F(\cg_t(\tau))]\ =\ {\bf E}[\left(
   \al_0(\nabla,\tau)+\hf\beta^2(\nabla,\tau)\right)F(\cg_t(\tau))]
\label{dtE}
\een
with
\ben
  \al_0(\nabla,\tau)\ =\ \sum_{n}a_n(\tau)\nabla_n,\ \ \ \ \ \ \ \ \ \ 
   \beta(\nabla,\tau)\ =\ \sum_{n}b_n(\tau)\nabla_n
\label{al0nabla}
\een

We now return to (\ref{0}) and shall consider the situation
where it corresponds to a null vector at level two 
in the indecomposable highest-weight module generated
from $\ket{\D+\theta}$ \cite{Flo-9707,MRS}. 
Such a logarithmic null vector, $\ket{\chi(c,\D),\theta}$, at level two
(or one, see below) may be characterized by the vanishing conditions
\ben
 L_1\ket{\chi(c,\D),\theta}\ =\ L_2\ket{\chi(c,\D),\theta}\ =\ 0
\label{LL0}
\een 
These conditions turn out to be too restrictive
for the characterization of 
logarithmic null vectors at levels higher than two \cite{Flo-9707,MRS},
but suffice for our level-two purposes.
Imposing $L_{-1}\ket{\D+\theta}=0$ implies 
$0=L_1L_{-1}\ket{\D+\theta}=2L_{0}\ket{\D+\theta}=2(\D+\theta)\ket{\D+\theta}$,
and since we naturally require $\D$
to be real (or at least complex), the state $L_{-1}\ket{\D+\theta}$
is not a logarithmic null vector. From (\ref{LL0}) it then follows
that a logarithmic null vector at level two is of the form
\ben
 \ket{\chi(c,\D),\theta}\ =\ \left(-2L_{-2}+\g L_{-1}^2\right)\ket{\D+\theta}
\label{-2L}
\een
where
\bea
 \g&=&\frac{3}{2\D+1+2\theta}\ =\ \frac{3}{2\D+1}
   -\frac{6\theta}{(2\D+1)^2}\nn
 c&=&(6\g-8)(\D+\theta)\ =\ 
  \frac{2\D(5-8\D)}{2\D+1}-\frac{32(\D-\frac{1}{4})(\D+\frac{5}{4})\theta}{
    (2\D+1)^2}
\label{gc}
\eea
With $\D$ and $c$ independent of $\theta$ we thus conclude that
a logarithmic null vector at level two exists exactly when
$(c,\D)$ is $(1,1/4)$ or $(25,-5/4)$,  
in accordance with \cite{Flo-9707,MRS}.
A comparison of this with (\ref{0}) and (\ref{SLEG}) tells us
that a logarithmic null vector is constructed provided that $\tau=\theta$ and
\ben
 k(\theta)\ =\ \frac{6}{2(\D+\theta)+1}\ =\ \frac{6}{2\D+1}
  -\frac{12\theta}{(2\D+1)^2}
\label{kt}
\een
We immediately recognize the announced need for a $\theta$-dependent
(or $\tau$-dependent) walk on the Virasoro group (\ref{cg}). 
It follows from (\ref{kappa}) and (\ref{kt})
that the bulk part for $\D=1/4$ corresponds to SLE$_{\kappa=4}$.
The other conformal weight, $\D=-5/4$, admitting a logarithmic null vector
at level two would result in a negative $\kappa$ and is therefore 
not related immediately to ordinary SLE.

\section{Conclusion}

We have discussed how certain stochastic evolutions may be linked
to LCFT. The method and results extend the work of Bauer and Bernard
\cite{BB} and are in the vein of the graded extensions found in \cite{Ras,NR}.
The connection is established through the introduction of nilpotent
parameters which allow one to treat the logarithmic fields in a unified way
\cite{Flo-9707,MRS}, and to extend the realm of the conformal maps involved.
Emphasis has been put on a simple scenario
corresponding to a straightforward extension of SLE.
The new system treats the L\"owner differential equation as 
belonging to a pair of coupled differential equations.
Ordinary SLE is recovered by disregarding the second equation.
The general link is quite formal but can be made more direct by
relating the representation theory of LCFT to entities conserved
in mean under the stochastic process. This has been discussed
explicitly in the scenario just mentioned. In a particularly interesting case, 
the bulk part then corresponds to SLE$_4$.
\vskip.5cm
\noindent{\em Acknowledgements}
\vskip.1cm
\noindent The author thanks M. Flohr, P. Mathieu and O. Schramm
for comments.
\vskip.5cm
\noindent{\em Note added}
\vskip.1cm
\noindent After completion of the present work, the paper \cite{MRR} 
has appeared. It also concerns the formal link between LCFT
and SLE, and contains some of the results derived above, notably the
case resembling SLE as based on (\ref{dhdh}) and (\ref{-2L}).

\end{document}